\documentclass{PoS}

\usepackage[utf8]{inputenc}
\usepackage{graphicx}
\usepackage{amsmath,amssymb}

\def\Tr{{\rm Tr}\,}

\title{Large-$N$ reduction with two adjoint Dirac fermions}
\ShortTitle{Large-$N$ reduction with two adjoint Dirac fermions}

\author{Barak Bringoltz,$^{ab}$ \speaker{Mateusz Koren}$^{cb}$ and Stephen
R.~Sharpe$^{b}$\\
\llap{$^a$}IIAR -- the Israeli Institute for Advanced Research,\\
Rehovot, Israel\\
\llap{$^b$}Department of Physics, University of Washington,\\
Seattle, WA 98195-1560, USA\\
\llap{$^c$}M. Smoluchowski Institute of Physics, Jagiellonian University,\\
Reymonta 4, 30-059 Cracow, Poland\\
E-mail: \email{barak.bringoltz@gmail.com},
\email{mateusz.koren@uj.edu.pl}, \email{sharpe@phys.washington.edu}}

\abstract{We study the single site $SU(N)$ lattice gauge theory with $N_f=2$
adjoint Wilson fermions for values of $N$ up to 53. We determine the phase
diagram of the theory as a function of the hopping parameter $\kappa$ and the
inverse 't Hooft coupling $b$, searching for the region in which the
$\mathbb{Z}_N^4$ center symmetry is unbroken. In this region the theory is
equivalent to the infinite volume theory when $N\to\infty$. We find a region of
values of $\kappa$ on both sides of $\kappa_c$ for which the symmetry is
unbroken, including both light physical quarks and masses $\sim {\cal O}(1/a)$.
This is surrounded by a region with a complicated sequence of partially broken
phases. We calculate Wilson loop expectation values and find that using $N \leq
53$ it is possible to extract the heavy-quark potential at small distances (1-3
links) but not at longer distances. For this, larger values of $N$, or lattices
with more sites, are needed.}
        
\FullConference{The XXIX International Symposium on Lattice Field Theory - Lattice 2011\\
July 10-16, 2011\\
Squaw Valley, Lake Tahoe, California}

\begin{document}

\section{Introduction}

A key property of large-$N$ lattice gauge theories is volume
reduction, first introduced by Eguchi and Kawai (EK) \cite{EK82}.
They showed that, under some assumptions, Yang-Mills theory on an infinite
lattice satisfies the same Dyson-Schwinger equations 
as the theory reduced to a single space-time point.
The crucial assumption was that the center symmetry of the action cannot be 
spontaneously broken. This assumption was soon shown to be invalid~\cite{BHN82}.
Over the years, many alternatives have been proposed to restore EK equivalence,
but until recently, none has been fully satisfactory.\footnote{%
For an overview see, e.g., Ref.~\cite{MU11}.
One highly successful alternative approach 
due to Narayanan and Neuberger is the reduction to
a small volume of size $\sim (1 {\rm fm})^4$ rather than
to a single site.}

The recent revival of this topic was triggered by the work of
Refs.~\cite{HN02, KUY04} which showed that
volume reduction was an example of broad
class of large-$N$ orbifold equivalences, holding
both in the continuum and on the lattice. If
one can construct orbifold projections between the parent and
daughter theories (in our case the large-volume and single-site theories,
respectively) then the two become equivalent for $N\to\infty$,
assuming certain conditions hold. 
%
%
The most non-trivial of these conditions is, as above,
the preservation of the center symmetry. The authors of
Refs.~\cite{KUY04, KUY07} proposed two possible fixes that might stabilize the
ground state to preserve the center symmetry. 
They calculated the perturbative potential
for the eigenvalues of the Polyakov loop 
and showed that it becomes repulsive
if one adds $N_f>1/2$ massless Dirac
fermions in the adjoint representation (with periodic boundary
conditions).\footnote{The other possibility is the trace-deformed EK model,
which, however, becomes very complex when one wants to compactify 
more than one space-time dimension (see e.g. Ref.~\cite{HV11}).}
In perturbation theory this leads to a center-symmetric ground state, with
eigenvalues spread uniformly around the unit circle.
Apart from being interesting in their own right,
theories with $N_f$ adjoint fermions are connected by a chain
of large-$N$ equivalences to QCD with $2N_f$ fundamental flavours.
This opens up the possibility of using single-site
simulations to give non-perturbative
insight into the strong interactions (up to $1/N$ corrections).

In the following we focus on a single-site $SU(N)$ lattice model with
adjoint Wilson fermions--the Adjoint Eguchi-Kawai or AEK model. 
The partition function is
\begin{equation}
	\mathcal{Z}=\int D[U,\psi,\bar\psi]\,e^{(S_{\rm gauge} +
  \sum_{j=1}^{N_f} \bar \psi_j \, D_{\rm W} \, \psi_j)},
\end{equation}
where $S_{\rm gauge}$ is the single-site equivalent of the Wilson action:
\begin{equation}
S_{\rm gauge} = 2 N b\, \sum_{\mu<\nu} {\rm Re}\Tr U_\mu U_\nu U^\dagger_\mu
U^\dagger_\nu
\end{equation}
($b=\tfrac{1}{g^2 N}$ being the inverse 't Hooft coupling)
and $D_W$ is the single-site Wilson Dirac operator:
\begin{equation}
D_W = 1 - \kappa \sum_{\mu=1}^4 \left[ \left( 1 - \gamma_\mu\right) U^{\rm
adj}_\mu + \left(1 + \gamma_\mu\right)U^{\dag {\rm adj}}_{\mu}\right].
\end{equation}

Using massless fermions is crucial in the perturbative analysis of
Ref.~\cite{KUY07} but a lattice simulation by two of us of the
$N_f=1$ AEK model found that, for $N\leq15$,
center symmetry is stabilized not only by light fermions
but also by fermions with mass $~1/a$~\cite{BS09}.
This is much larger than the expected upper limit of $~1/(aN)$~\cite{HM09,BB09}. 
If this continues to hold for $N\to\infty$,
then the $N_f=1$ AEK model provides a realization of the
original Eguchi-Kawai idea. 
This result was confirmed, and given a semi-analytic
understanding (using arguments going beyond perturbation theory) 
in Ref.~\cite{AHUY}. 
Simulations using massless overlap fermions have also found evidence
for reduction~\cite{Hietanen}.
There has also been work on the $N_f=2$ theory also suggesting
that reduction holds~\cite{Catterall}.

\section{Phase diagram of $N_f=2$ Adjoint Eguchi-Kawai model}

From now on we focus on the AEK model with $N_f=2$.
In addition to the above-mentioned connection to QCD with 4 fundamental
flavors, the large volume theory to which it might be equivalent
is of phenomenological interest as a theory which might
exhibit conformal, or near-conformal, behavior in the IR.
The $N=2$ version of this theory has been extensively studied (see
Ref.~\cite{DSS11} and references therein).

We use the Hybrid Monte Carlo algorithm 
adapted to work on a single-site lattice
(for details see Ref.~\cite{BKS11}). 
The CPU time scales approximately as
$N^4$ for heavy fermions and $N^{4.5}$ for light fermions.~\footnote{%
The contributions to the scaling are
$N^3$ for fundamental link matrix multiplication 
(adjoint link matrices are never explicitly constructed
in the updating procedure),
$\sim N$ for the growth in the number of
molecular dynamics steps for constant acceptance,
and, for light fermions, $\sim N^{0.5}$ for the
increase in the number of CG iterations.}
This is a significant improvement over 
the $\approx N^8$ growth for the Metropolis algorithm used in
Ref.~\cite{BS09},
and allows us both to reach higher values of $N$ and reduce
statistical errors.

To analyze the phase diagram we have performed scans in the $\kappa-b$ plane
using values of $N$ up to 30. 
In addition, we have done several high-statistics
runs at selected points in parameter space using $N\leq53$. We have
performed simulations with values of $b$ up to $b=200$ but we have
mostly focused on the region $b\in[0,1]$. 
In particular, the two values we have looked at with most care are:
\begin{itemize}
\item{$b=0.35$, which is close to the value used in typical lattice simulations
(for $N=3$ we have $\beta=6/g^2=6.3$),}
\item {$b=1.0$, which is a fairly weak coupling ($\beta=18$ for $N=3$) at
which one should be able to compare the results to perturbation theory.}
\end{itemize}
 
Apart from analyzing the plaquette, we have used two types of observables 
to study center symmetry breaking:
\begin{enumerate}
  \item {General ``open loops'':
\begin{equation}
K_{n}\equiv \frac1{N}\Tr\, U^{n_1}_1\, U^{n_2}_2\, U^{n_3}_3\, U^{n_4}_4, 
\quad {\rm with}\ \ n_\mu =0,\pm 1, \pm 2, \dots
\end{equation}
where $U^{-n} \equiv U^{\dag n}$. These are the general order parameters
for the breaking of the center symmetry. The simplest example of such loops are
the Polyakov loops $P_\mu = \Tr U_\mu$.}
\item{Eigenvalues of link matrices: the single-site gauge
transformation $U_\mu \to \Omega U_\mu \Omega^\dagger$ obviously leaves the set
of eigenvalues unchanged. In the center symmetric-phase one expects the
distribution of phases of the eigenvalues to be invariant under translations by
$2\pi n/N$.

At weak coupling one also expects that the links are close to being
simultaneously diagonalizable. We thus apply the gauge transformation that
diagonalizes one link and look at the elements of the other links. 
Indeed, we
find that for $b\gtrsim1$ the diagonal elements of the links
dominate in this basis while as we move to stronger couplings the off-diagonal
elements grow the links become increasingly non-commutative.}
\end{enumerate}

\begin{figure}[btp!] 
\centering
\includegraphics[width=9cm]{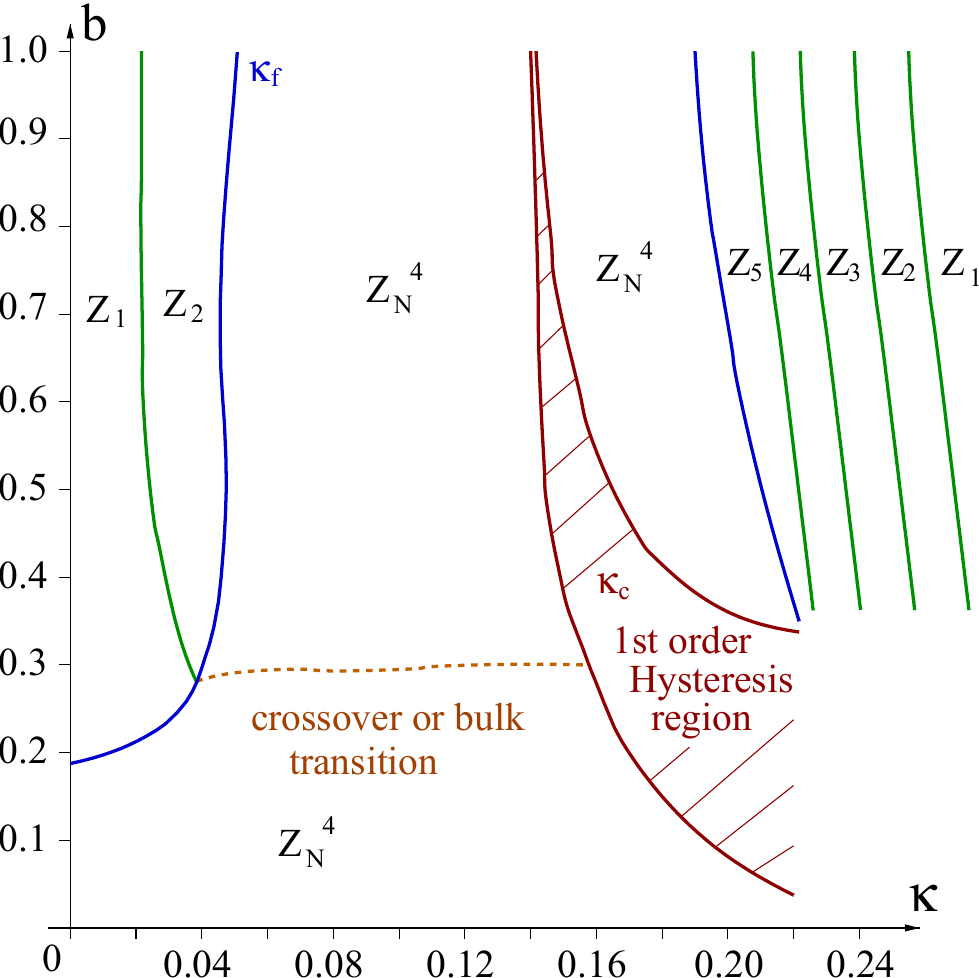} 
\caption{Sketch of the
phase diagram for the $N_f=2$ AEK model. The positions of phase boundaries are
approximate, and depend somewhat on $N$. The funnel gets narrower very slowly
when moving to larger values of $b$. We have also found evidence for a
$\mathbb{Z}_3$ phase on the left-hand side of the funnel for $b\gtrsim5$.}
\label{fig:phase_diag}
\end{figure}

The resulting phase diagram is pictured in Figure \ref{fig:phase_diag}. Note
that this diagram is qualitatively similar to the one from Ref.~\cite{BS09}
for the $N_f=1$ AEK model. The $\kappa=0$ line corresponds to the
original Eguchi-Kawai model with broken center symmetry at weak coupling. 
As the fermions become lighter a ``funnel'' of center-symmetric phase 
appears on both sides of the critical region. 
This is the region in which volume reduction holds. 
Around it there are regions in which the $\mathbb{Z}^4_N$ center symmetry
is broken---moving away from the funnel
the breaking is to increasingly small subgroups until
we reach a completely broken phase.
We label the phases by the approximate remnant of the center
symmetry. In particular, a ``$\mathbb{Z}_i$'' phase has $i$ clumps
in the histograms of the phases of link eigenvalues.
In the broken phases there are strong correlations
between links in all directions---this is why we label these phases with
$\mathbb{Z}_i$ instead of $\mathbb{Z}^4_i$. For a more detailed analysis of the
phase structure, see Ref.~\cite{BKS11}.

An important issue is how the width of the funnel depends on
$N$ and $b$. As shown in Fig.~\ref{fig:phase_diag}, we denote
the lower limit of the funnel by $\kappa_f$,
and the corresponding quark mass by $m_f=1/(2\kappa_f)-1/(2\kappa_c)$).
We first discuss the $b$ dependence.
The analysis of Ref.~\cite{AHUY}
predicts that at weak coupling the center-symmetric and completely broken
phases are separated by many phases with partial breakings 
(which is in agreement with our results) 
and that the funnel closes as $m_f\sim b^{-1/4}$ for $b\to\infty$.
We have performed simulations for $N=10$ up to
$b=200$ and our results are consistent with this form.
It is important to note, however, that, although the
funnel closes, it does so in such a way that
quarks of arbitrarily high {\rm physical} mass,
$m_f =a m_{\rm phys}$, remain inside the funnel for $b\to\infty$.

Using runs up to $N=30$, we find that $\kappa_f$ increases with $N$, 
so that the width of the funnel shrinks.
It is thus very important to go to even higher $N$ in order
to perform a reliable extrapolation of $\kappa_f$ to $N\to\infty$,
and thus to establish whether the funnel (at fixed $b$)
has finite width at $N=\infty$ or not.
At this stage, the strongest evidence we have for a finite
width is obtained by studying the large-$N$ limit of the
plaquette within the putative funnel region.
We have done this extrapolation at several values of $\kappa$
for $b=0.35$ and $1.0$, using $N$ up to 53.
We find that the extrapolated plaquette for $\kappa$ away from $\kappa_c$
is almost independent of $\kappa$, and furthermore lies close
to the value obtained in the large-$N$ pure gauge theory~\cite{BKS11}.
This is what we would expect if reduction holds, because the
heavy quarks lead only to a small renormalization of the plaquette.

Another interesting aspect of the phase diagram is the order of the critical
line. Large volume $SU(2)$ simulations with adjoint fermions
find a second order transition line emanating from $b=\infty$, $\kappa=1/8$
which becomes first order below $b\approx 1/4$. This structure
is related to the claimed presence of an IR fixed point.
Although $N=2$ is far from the $N=\infty$ limit, it is expected
for the adjoint theory that its properties should depend only
weakly on $N$.

\begin{figure}[btp!] \centering 
\includegraphics[width=10.5cm]{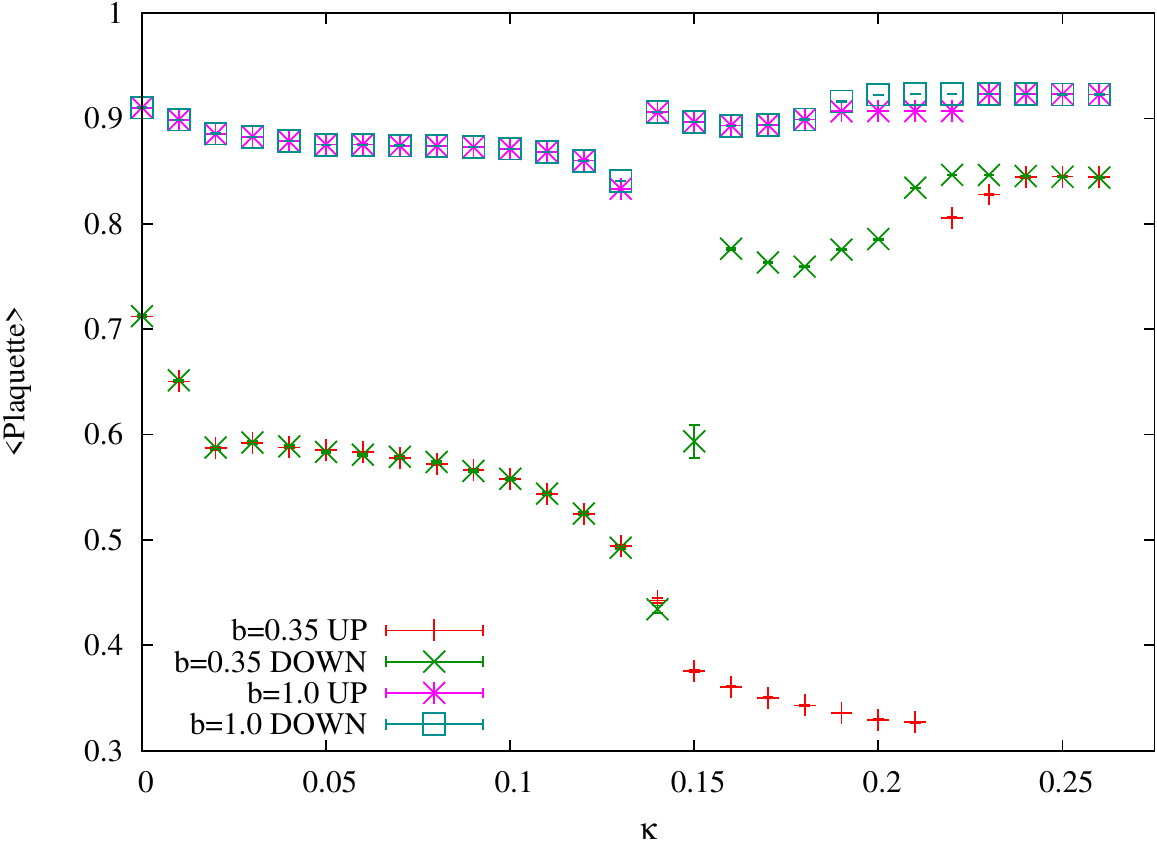}
\caption{Scans of the average plaquette for $N=23$ 
at $b=0.35$ and $1$.}
\label{fig:plaq}
\end{figure}

In fact, we find a very different phase diagram, in which
there is a clear first-order transition for $b$ up to
at least $1$.
This is illustrated by the results
shown in Fig.~\ref{fig:plaq}. 
The discontinuity in the plaquette drops as $N$ increases,
but, for $b\leq 1$, we find that it
remains non-zero in the $N\to\infty$ limit.
For larger values of $b$ our results are less definitive.
Although we still see a discontinuity, its magnitude
decreases rapidly with increasing $b$,
and it is difficult to determine whether it remains as $N\to\infty$.
Nevertheless, it is clear that the first-order transition
extends to much larger $b$ than in the large-volume $N=2$
simulations. The simplest interpretation of our results is
that the theory is confining in the IR, with lattice artifacts
leading to a first-order transition at $\kappa_c$ which
extends to $b\to\infty$. 

\section{Towards physical quantities}

The ultimate goal of using volume-reduced simulations is to calculate physical
quantities that one can apply to large volume theories.
One such quantity
is the heavy quark potential,
which can be obtained using rectangular Wilson loops. In the 
single-site theory these become
\begin{equation}
W(L_1,L_2) = \frac1{12}\sum_{\mu\ne\nu} \langle \tfrac1N \Tr U_\mu^{L_1}
U_\nu^{L_2} U_\mu^{\dagger L_1}U_\nu^{\dagger L_2}\rangle\,.
\end{equation}
For large $L_2$ we expect
$W(L_1,L_2)\sim e^{-V(L_1)L_2}$ but at finite $N$ 
we must keep $L_j < N$ to avoid finite $N$ effects.

\begin{figure}[btp!] \centering 
\includegraphics[width=10.5cm]{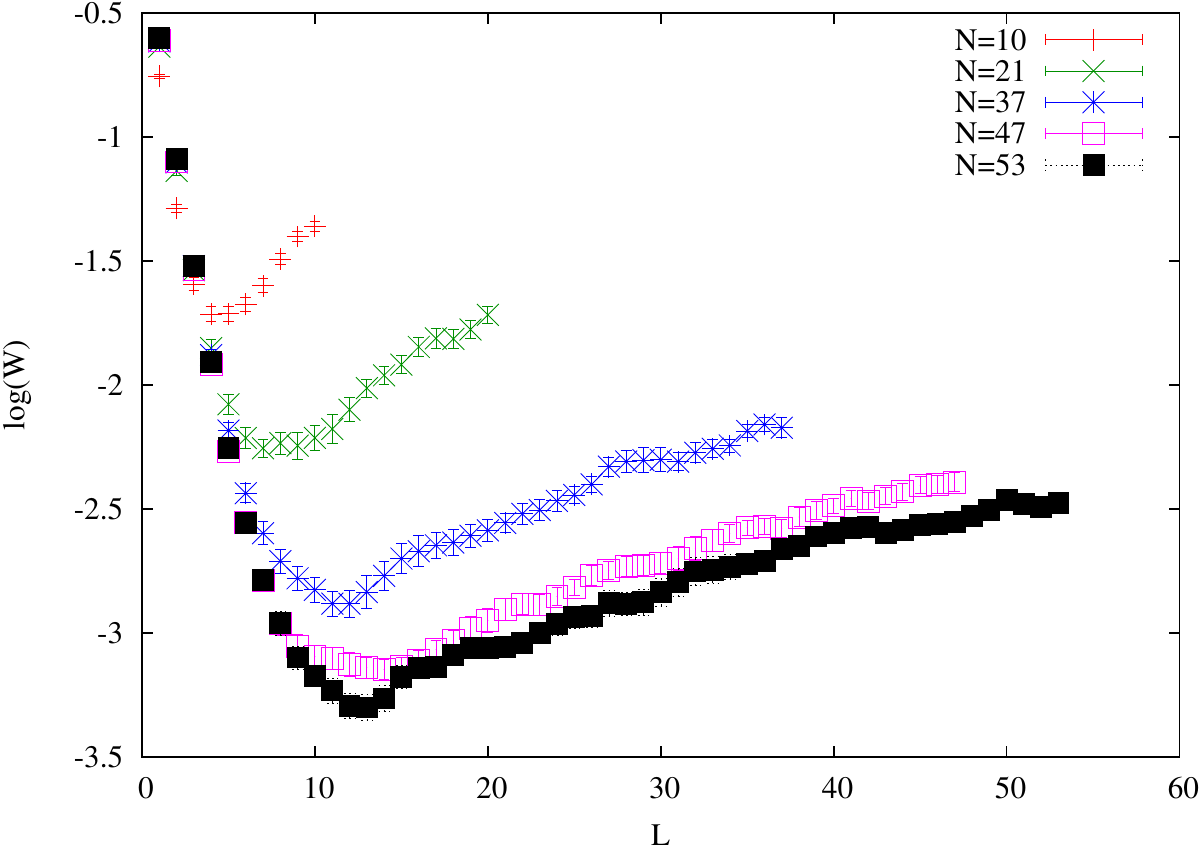}
\caption{Log-linear plot of $1\times L$ Wilson loop versus $L$ for $L\le N$.}
\label{fig:wloops}
\end{figure}

Figure~\ref{fig:wloops} shows a log-linear plot of $1\times L$ Wilson loops
for several values of $N$. We see convergence to a common envelope 
for small $L$ followed by a slow rise.
The latter is a finite-$N$ effect---as $N\to\infty$ the 
exponentially falling envelope will extend to $L=\infty$.
We stress that this finite-$N$ effect is statistically 
significant.
From the minima of the curves, one
can estimate how $L_{\rm max}$ scales with $N$.
The dependence is approximately logarithmic,
so that calculation of the potential 
poses a significant numerical challenge.
The problem becomes 
worse as we move to larger separations and
we are only able to see  convergence to 
the linear envelope (and thus extract the
potential) for distances of 1-3 links. 
For larger distances one needs to either
use larger values of $N$ or move to larger
lattices (e.g. $2^4$).

\section{Summary \& outlook}

The single-site lattice gauge theory 
with 2 flavours of adjoint Wilson-Dirac fermions
exhibits a rich phase structure,
one that is qualitatively explained by the semi-analytic model of
Ref.~\cite{AHUY}. Our most important result is that, 
for $N$ up to 53, we find a
broad funnel of center-symmetric phase where large-$N$ reduction holds. It
extends from strong to very weak coupling and contains fermions with
masses up to $\sim1/a$. 
The $N\to\infty$ extrapolations of the plaquette suggest
that the funnel does not close in the large-$N$ limit but we cannot completely
rule out this possibility at present.
A detailed analysis using even higher $N$ is desirable.

We have also analyzed the order of the critical line $\kappa_c$, finding
a first-order phase transition for all values $b$. This is different from
the large volume $SU(2)$ simulations and favors a confining scenario
over the conformal one for the large-$N$ theory.

To use the large-$N$ volume reduction as a practical tool one needs to be able
calculate long-distance observables. We have presented a calculation of the
heavy-quark potential and showed that $1/N$ effects limit the precision that one
can achieve.
 
This analysis can be extended by calculating more physical quantities 
such as particle masses, as well as extending the
range over which the heavy-quark potential can be extracted.
For the latter it is desirable to reduce the ``$1/N$ noise'' e.g. by using
slightly larger lattices or using twisted model as in Ref. \cite{AHUY}. 
In addition,
making a careful analysis of the vicinity of the critical line, perhaps using
improved actions, would be of considerable interest.

\section*{Acknowledgments}

This work was supported in part by the U.S. DOE Grant No. DE-FG02-96ER40956, 
and by the Foundation for Polish Science MPD Programme co-financed by the 
European Regional Development Fund, agreement no. MPD/2009/6.

\end{document}